\documentstyle[twocolumn,aas2pp4]{article}
\def\stacksymbols #1#2#3#4{\def\theguybelow{#2}
	\def\verticalposition{\lower#3pt}
	\def\spacingwithinsymbol{\baselineskip0pt\lineskip#4pt}
	\mathrel{\mathpalette\intermediary#1}}
\def\intermediary #1#2{\verticalposition\vbox{\spacingwithinsymbol
	\everycr={}\tabskip0pt
	\halign{$\mathsurround0pt#1\hfil##\hfil$\crcr#2\crcr
		\theguybelow\crcr}}}
\def\lta{\stacksymbols{<}{\sim}{2.5}{.2}}
\def\gta{\stacksymbols{>}{\sim}{3}{.5}}

\begin{document}
\title{RECENT X-RAY OBSERVATIONS AND 
THE EVOLUTION OF HOT GAS IN ELLIPTICAL GALAXIES:
EVIDENCE FOR CIRCUMGALACTIC GAS$^1$}

\author{Fabrizio Brighenti$^{2,3}$ and William G. Mathews$^3$}

\affil{$^2$Dipartimento di Astronomia,
Universit\`a di Bologna,
via Zamboni 33,
Bologna 40126, Italy\\
brighenti@astbo3.bo.astro.it}

\affil{$^3$University of California Observatories/Lick Observatory,
Board of Studies in Astronomy and Astrophysics,
University of California, Santa Cruz, CA 95064\\
mathews@lick.ucsc.edu}

\begin{abstract}
The radial variation of hot gas density and temperature 
in bright elliptical galaxies 
provided by x-ray observations can be used to 
accurately determine the radial distribution of 
total galactic mass using the condition of hydrostatic 
equilibrium.
However, we set out here to solve the inverse problem.
Starting with the known distributions of total mass 
and mass-losing stars in a 
specific large elliptical, NGC 4472, 
we attempt to 
solve the gas dynamical equations to recover the currently 
observed radial distribution of density and temperature 
in the hot interstellar gas.
The galaxy is assumed to be initially gas-free as a 
result of early SNII-driven galactic winds.
In seeking this agreement we consider a variety of 
assumptions for the evolution of the hot interstellar 
gas: mass dropout, variation of stellar mass loss with 
galactic radius, variation of the bolometric radiative 
cooling rate with galactic radius, and several 
supernova rates.
After evolving for a Hubble time, none of these models 
accounts for several well-observed properties of the interstellar
gas: gas temperatures that significantly 
exceed the mean stellar temperatures, 
positive temperature gradients within a few effective radii,
shallow density gradients, and large interstellar gas masses.

However, all of these discrepancies are lessened 
or disappear if
large masses of hot gas are assumed to be present 
in the outer galactic potential at early times in 
galactic history.
Most of the interstellar mass that contributes to
cooling flows in galaxies like NGC 4472 has not come from
mass lost by galactic stars since $\sim 1$ Gyr.
The sustained inflow of hot gas from this circumgalactic 
environment may help to resolve other long-standing 
problems that 
have beset models of galactic cooling flows: 
the wide range of $L_x/L_B$
for fixed $L_B$; the variable and often low iron abundances 
observed in ellipticals, and 
the failure (so far) to observe pronounced 
rotational flattening 
in x-ray images of slowly rotating giant ellipticals.

The characteristic hot gas temperature profile observed 
in many bright ellipticals has a maximum at $\sim 3~r_e$.
This can be understood as the mixing of gas ejected from stars
with old circumgalactic gas flowing in from the halo.
Since the circumgalactic gas is hotter than the stars,
mass dropout is not needed to flatten the gas density gradient.
Moreover, the agreement of the stellar mass in NGC 4472 
(and NGC 4649) 
within $r_e$ determined from stellar dynamics 
with the total mass determined by hot gas 
hydrostatic equilibrium is an additional argument against
mass dropout.

\end{abstract}


\section{INTRODUCTION}

As the quality of x-ray observations of elliptical galaxies has 
increased, the modest agreement with simple theoretical models 
has diverged.
In bright ellipticals the x-ray emission is dominated by 
thermal radiation from optically thin interstellar gas.
The radial variation of density and temperature in the hot
gas, $\rho(r)$ and $T(r)$, 
can be determined from observations of the projected
X-ray surface brightness and spectrum.
But we show here that neither $\rho(r)$ nor $T(r)$ can be 
reproduced with gas dynamical calculations based 
on variations of the standard assumptions of galactic 
cooling flow theory.

In the usual theoretical approach the equations of gas dynamics 
are solved with source terms describing the rate that 
new gas flows 
into the interstellar gas, ejected by red giants in 
an old stellar population.
The thermal energy of the interstellar gas is influenced by 
several effects:
(i) the (kinetic) 
energy of newly injected gas that it inherits from the orbital
motion of the parent stars,
(ii) additional heating by Type Ia supernovae,
and, most importantly, (iii) compressive heating in the 
galactic gravitational potential.
The final gas temperature, $T \sim 10^7$ K, 
is comparable to the virial temperature of the galactic potential.
Heating by supernovae cannot be too large otherwise the 
iron abundance in the hot interstellar gas would exceed 
observational limits 
or the heated gas would flow out of the galaxy as a wind,
reducing the total x-ray luminosity $L_x$ below the threshold 
of detectability.
Conversely, 
the interstellar gas cannot be free-falling rapidly 
in the galactic potential since 
the mass loss rate from evolving stars is insufficient 
to resupply the observed mass of interstellar gas in the 
freefall time for large ellipticals, $\lta 10^9$ yrs.
For these reasons it follows that the hot interstellar gas 
is flowing very subsonically and the condition of 
hydrostatic equilibrium is satisfied 
to an excellent approximation at 
most galactic radii.

The main properties of gas in x-ray luminous 
ellipticals have been confirmed with evolutionary 
gas dynamical calculations, but many important details remain 
unexplained.
Some inconsistencies between theory and observations
have been known for some time.
For example: the wide 
scatter in x-ray luminosity $L_x$ for galaxies of similar optical luminosity
$L_B$ (Eskridge et al. 1995) and the often surprisingly low 
(and varied) mean iron abundance in the hot gas compared to that 
observed in the stars (Arimoto et al. 1997).
It has also been recognized for many years that gas density 
slopes $d \rho /d r$ (derived from surface brightness observations) 
are less steep than those  
predicted by the most straightforward theoretical models. 
Recently, however, new inconsistencies have emerged.
New observations of the gas temperature using ROSAT PSPC and ASCA
(e.g. Davis and White 1996) indicate that the average 
gas temperature in massive ellipticals is $\sim 1.5 - 2$ times larger 
than the equivalent orbital temperature of the stars 
$T_* = (\mu m_p / k) \sigma_*^2$ where $\mu = 0.63$ is the 
mean molecular weight and $m_p$ is the proton mass.
The gas is much hotter than the stars.
In addition gas temperature profiles $T(r)$ in most or all of the 
massive ellipticals studied so far show an increase 
by $30 - 40$ percent from the central values to 
about three effective radii where the temperature flattens or slowly 
decreases toward larger radii (Jones et al. 1997; David et al. 1994;
Trinchieri et al. 1994; Mushotzky et al. 1994; 
Kim \& Fabbiano 1995; Irwin \& Sarazin 
1996; Trinchieri et al. 1997).
A plot showing these results can be found in 
Brighenti \& Mathews (1997a).
We show below that this peculiar temperature variation 
does not arise naturally in evolutionary flows based on 
gas lost from galactic stars. 
Finally, since all elliptical galaxies rotate at some significant 
level, their x-ray contours are expected to be significantly 
elongated perpendicular to the axis of rotation out to 
an effective radius or more (Brighenti \&
Mathews 1996; 1997b), yet most observed x-ray contours appear to be 
approximately circular!

In view of the serious nature of these observational inconsistencies 
with prevailing theory, in this paper we restrict our 
exploration of alternate theoretical models 
to spherical geometry. 
Our approach is to begin with a simple 
gas-dynamical model for the evolution 
of hot gas in a realistic galactic potential 
based on NGC 4472. 
We show that the final gas density and temperature variation in 
this model disagrees with observations of NGC 4472 in 
a variety of important ways.
Then we vary the model 
parameters and assumptions to seek improvements, but instead 
only illustrate how insensitive the model is even to rather drastic 
alterations of the parameters.
Finally, we show that the disagreement between the 
properties of interstellar gas in our gas dynamical model 
and those observed in NGC 4472 can be alleviated if large amounts 
of additional gas are supplied 
near the beginning of the galactic evolution. 
The same desirable 
result may be achieved by adding gas to the galaxy more slowly 
over many Gyrs, but we do not explore this possibility here.
The important point is that elliptical galaxies can retain
halos filled with very old hot gas.
This gas may be a relic of an earlier 
time when these massive ellipticals
were formed near the central focus of small groups.
For some large ellipticals in Virgo this ancient 
gas has survived after the galaxy has entered 
the environment of this rich cluster.

\section{NGC 4472 AND THE STANDARD ISM MODEL}

Recently we have reviewed the 
x-ray observations of three bright nearby ellipticals:
NGC 4472, 4636, and 4649.
From the known gas density and temperature profiles 
in these galaxies and their optical properties, 
we determined the variation of total
and stellar mass densities as functions of galactic radius 
assuming that the hot interstellar gas is in 
hydrostatic equilibrium (Brighenti \& Mathews 1997a).
For two of these galaxies, NGC 4472 and 4649, the hot interstellar 
gas is in equilibrium with the potential of the stars 
for $r \lta r_e$, the effective radius, 
but the potential of the 
massive dark halos dominates beyond $r_e$. 
For these two galaxies 
the agreement of the total mass and density with that 
of the stars alone 
over $0.1 \lta r/r_e \lta 1$ 
verifies that the mass to light ratio determined from the stars 
and the gas temperature determined from x-ray spectra 
are accurate and that hydrostatic equilibrium maintains.
We can therefore be confident that the underlying mass structure
of these galaxies is well understood.

\subsection{Properties of NGC 4472}

For a realistic galactic model in which to test various
gas dynamical models described below, 
we use the total mass $M_{tot}(r) = M_{dark}(r) + M_*(r)$
of NGC 4472 found by assuming hydrostatic equilibrium
using gas temperatures and densities based on 
x-ray observations. 
The variation of gas density $n(r)$
for NGC 4472
plotted in Figure 1a combines results from ROSAT HRI and PSPC
observations of Irwin \& Sarazin (1996)
and {\it Einstein} HRI observations of Trinchieri, Fabbiano
\& Canizares (1986).
The analytic fit to the gas density 
$n(r) = \rho(r)/ 1.25 m_p$ in Figure 1a is sum of functions
$n(r) = \Sigma n_i(r)$ where $n_i(r) = n_{o}(i) [1 +
[r/r_{o}(i)]^{p(i)}]^{-1}$.
For 4472 we used three such functions 
with the following parameters:
$n_{o}(i) = 0.095$, 0.00597, -.0004 cm$^{-3}$;
$r_{o}(i) = 0.107$, 0.95, 10. in units of $r_e$;
$p(i) = 2.0$, 1.14, 1.19.
The gas temperature profile for NGC 4472
in Figure 1a is fit with a function of the form
$T(r) = 2 T_m [ r_m/(r + r_{ot}) + (r/r_m)^q]^{-1}$ with
parameters $T_m = 0.75 \times 10^7$ K,
$r_m = 0.5r_e$, $r_{ot} = 0.75r_e$ and $q = 0$.
Although the observed temperature is an average along the line
of sight at projected radius $R$ weighted
by the local emissivity $\sim \rho^2$, we assume that
the temperature at physical radius $r$ is the same, i.e.
$T(r) \approx T(R)$.
This approximation is appropriate because of
the steep decline of $\rho$ with galactic radius (Brighenti \&
Mathews 1997a). 

The total mass that confines the hot interstellar gas is found 
from the condition for hydrostatic equilibrium
$$M_{tot}(r) = - {k T(r) r \over G \mu m_p }
\left( {d \log \rho \over d \log r} + {d \log T \over d \log r}
\right).$$
$M_{tot}(r)$ is plotted in Figure 1b.
Also shown in Figure 1b 
is the de Vaucouleurs stellar mass distribution 
$M_*(r)$ determined from the optical data 
for NGC 4472: luminosity $L_B = 7.89 \times 10^{10} L_{B,\odot}$, 
distance $D = 17$ Mpc, effective radius $r_e = 8.57$ kpc,
stellar mass to light ratio 9.20 (van der Marel 1991),
and total stellar mass $M_{*t} = 7.26 \times 10^{11}$
$M_{\odot}$.
$M_{tot}$ and the local total density (Figure 1c) 
are dominated by the dark halo for $r \gta r_e$ but 
within $r_e$ the x-ray distribution is determined by the stellar 
potential alone.
As noted earlier, the total mass in $0.1 r_e \lta r \lta r_e$ 
is in excellent agreement with the expected stellar mass,
$M_{tot} \approx M_*$.
The galactic potential $GM_{tot}/r$ that we adopt for 
our dynamic models for gas flow in NGC 4472 is
based on $M_{tot}(r)$ shown in Figure 1b.

Iron abundance determinations for the hot gas 
in NGC 4472 vary with
method and instrument used: 
$z_{Fe} = 0.20$ (Serlemitsos et al. 1993 with BBXRT), 
$z_{Fe} = 1-2$ (Forman et al. 1993 with ROSAT), 
$z_{Fe} = 0.63$ (Awaki et al. 1994 with ASCA), and 
$z_{Fe} = 1.18$ (Buote \& Fabian 1997 with ASCA).
Some, but not all, of this variation is due to 
significant differences in the adopted abundance for the sun
or solar system (Ishimaru \& Arimoto 1997).
Arimoto et al. (1997) adopt a mean (emission-weighted) 
gas abundance 
$z_{Fe} = 0.33$ in solar units with
$z_{Fe \odot} = 4.68 \times 10^{-5}$.
In our models much of the interstellar iron comes
from stellar mass loss.
Gas ejected from stars at any radius is assumed to have the same
iron abundance as the local parent stars.
For the stellar iron abundance in NGC 4472 
we assume a solar value 
at the galactic center which 
decreases as $z_{*Fe} \propto r^{-0.3}$ 
far from the galactic core (Arimoto et al. 1997), i.e. 
$z_{*Fe}(r) = 1.0/[ 1 + (r/r_b)^2]^{0.15}$ where 
$r_b = 200$ pc is the core or ``break'' radius of NGC 4472
(Faber et al. 1997).
We have deliberately chosen a somewhat smaller 
central stellar iron abundance than the $z_{*Fe}(0) = 1.6$ 
adopted by Arimoto et al. (1997);
previously it had been thought that $z_{*Fe}$ followed 
the spectral features of Mg in the solar ratio, 
but it is now thought that Mg/Fe 
exceeds the solar ratio in bright ellipticals 
(Scott Trager, private communication).
Since all of these parameters are highly uncertain, 
the absolute iron abundances we calculate are very approximate. 
Our main interest here are the relative emissivity-weighted 
abundances $\langle z_{Fe}/z_{Fe \odot} \rangle$ 
among the various dynamical models.

\subsection{The Standard Model for the Hot Gas in NGC 4472: Model STD}

The standard gas dynamics equations that describe the evolution
of hot interstellar gas in ellipticals are:
$${ \partial \rho \over \partial t} 
+ {1 \over r^2} { \partial \over \partial r}
\left( r^2 \rho u \right) = \alpha \rho_*,$$
$$\rho \left( { \partial u \over \partial t}
+ u { \partial u \over \partial r} \right)
= - { \partial P \over \partial r}
- \rho {G M_{tot}(r) \over r^2} - \alpha \rho_* u,$$
and
$$ \rho {d \varepsilon \over dt } =
{P \over \rho} {d \rho \over d t}
- { \rho^2 \Lambda(T) \over m_p^2}
+ \alpha \rho_*
\left[ \varepsilon_o - \varepsilon - {P \over \rho}
+ {u^2 \over 2} \right].$$
Here $\varepsilon = 3 k T / 2 \mu m_p$ is the 
specific thermal energy.
Both stars and supernova contribute mass to the interstellar 
gas, $\alpha = \alpha_* + \alpha_{sn}$, 
but the contribution from supernovae $\alpha_{sn}$ is very small.
An old stellar population expels gas at a rate 
$\alpha_*(t) \rho_*$ gm cm$^{-3}$ s$^{-1}$ where 
$\alpha_*(t) = \alpha_*(t_n) (t/t_n)^{-1.3}$,
$t_n = 13$ Gyrs is the current time and 
$\alpha_*(t_n) = 5.4 \times 10^{-20}$ s$^{-1}$
(Mathews 1989).
The time variation of $\alpha_*(t)$ 
is surprisingly insensitive to the IMF 
assumed, but $\alpha_*(t_n)$ is only know to within 
a factor of $\sim 2$.
As described above, $M_{tot}(r)$ is known directly from 
the observed gas temperature and density in NGC 4472 and 
the stellar density $\rho_*(r)$ is based on a de Vaucouleurs 
profile normalized to values appropriate to NGC 4472.
The loss of thermal energy by optically thin thermal emission is
described by the cooling coefficient
$\Lambda (T)$ modeled after the results of 
Raymond, Cox \& Smith (1976).
The gas is assumed to be heated only by supernova of 
Type Ia since we assume that the early phase of 
galactic winds generated by Type II supernovae has 
already subsided at time $t = 1$ Gyr 
when we begin our calculations.
The mean gas injection energy is 
$\varepsilon_o = 3 k T_o /2 \mu m_p$ where
$T_o = (\alpha_* T_* + \alpha_{sn} T_{sn})/\alpha$.
The local stellar temperature $T_*(r)$ is found by 
solving the equation of stellar hydrodynamics 
(with isotropic velocity ellipsoids) 
in the total mass potential of NGC 4472. 
Supernova heating of the interstellar gas is described by 
$$\alpha_{sn} T_{sn} =
2.13 \times 10^{-8}~ {\rm SNu}(t)~ (E_{sn}/10^{51} {\rm ergs})~$$
$$h^{-1.7}~ (L_B/L_{B \odot})^{-0.35} ~~~ {\rm K}~{\rm s}^{-1}$$
where $h \equiv H/100 = 0.75$ is the reduced Hubble constant and we
adopt $E_{sn} = 10^{51}$ ergs as the typical energy
released in a supernova event.
This equation incorporates the mean stellar mass to light ratio 
for elliptical galaxies from van der Marel (1991),
$M_{*t} / L_B = 2.98 \times 10^{-3} (L_B/L_{B \odot})^{0.35} h^{1.7}$.
(The individually determined mass to light ratio 
for NGC 4472, $M_{*t} / L_B = 9.20$, is 
slightly lower than that of the average elliptical having the same 
$L_B$.)
For our galaxy model (based on NGC 4472) we take
$L_B = 7.89 \times 10^{10} L_{B \odot}$.
Although there is some information available about the 
current Type Ia supernova rate in ellipticals 
(Turatto, Cappellaro \& Benetti 1994), almost nothing 
is known about supernova rates in the distant past.
For this paper we adopt a power law to describe a 
varying rate of Type Ia supernovae:
$${\rm SNu}(t) = {\rm SNu}(t_n) (t / t_n)^{-p}$$
with $p = 1$ and 
${\rm SNu}(t_n) = 0.04$ SNu (1 SNu = 1 supernova every 
100 years in stars having luminosity $L_B = 10^{10} L_{B \odot}$).
We assume each supernova ejects 1.4 $M_{\odot}$ into the 
local interstellar gas and 
that half of this, 0.7 $M_{\odot}$, is iron.
The iron abundance currently observed in the hot gas 
provides a global constraint 
on the past supernova rate and our parameters are 
chosen with this constraint in mind.
Finally, the de Vaucouleurs model for the stellar component 
in our model galaxy extends out to radius 
$r_t = 100$ kpc where the stellar density is very low,
but the dark halo continues out to 470 kpc where its 
total mass is $M_{dark,t} = 2.7 \times 10^{13}$ $M_{\odot}$.
Since the gas is ejected from stars within 100 kpc, the 
much more extensive dark halo -- which may resemble those 
of central ellipticals
in groups -- is irrelevant to most of 
the following gas dynamical solutions.

The gas-dynamical equations are solved using a standard
one-dimensional second order Eulerian code. We use reflection 
boundary conditions at the origin and outflow boundary 
conditions at the outer boundary, $r = 470$ kpc.

\subsection{Comparison of Standard Model (STD) and Observations}

Along with the observations of NGC 4472 in Figure 2 we also 
show the current 
variation of density and temperature predicted by solving
the gas dynamical equations above.
These equations are solved beginning with 
no gas in the galaxy at $t = 1$ Gyr and carried forward to 
the current time, $t_n = 13$ Gyr.
This gas-free 
initial condition can be understood if SNII explosions 
had driven strong winds out of the galaxy at early times 
(Mathews 1989; avid et al. 1991; Loewenstein \& Mushotzky 1996).
The elliptical in our gas dynamical model 
is assumed to be isolated 
so that gas can flow out beyond the galaxy if it wishes.

The deficiencies of the model are apparent!
The predicted density gradient for this ``standard 
model'' (STD) shown in Figure 2 
is steeper than the observed density distribution 
throughout the entire galaxy. 
Traditionally this discrepancy has been recognized only 
in the central regions.
The projected x-ray surface brightness of the model galaxy 
$\Sigma_x(R)$ is also steeper than that observed for NGC 4472.
This density discrepancy still persists for models, not shown in 
Figure 2, in which the calculation is begun at 
$t = 0.1$ Gyr to increase (by $\sim 3$) the total amount 
of gas ejected from the stars.
Although the gas temperature predicted by our 
model is roughly comparable with 
observed values shown in Figure 2,
it is definitely too low at all radii by a factor of 
$\sim 2 - 3$ except perhaps 
within a few kpc from the center.
In addition, 
the shape of our computed temperature variation $T(r)$ 
does not have the characteristic
positive gradient $d T / d r$ out to $r \sim 3 r_e$ that is 
typical of most or all galactic cooling flows observed so far
(Brighenti \& Mathews 1997a).
For deep gravitational potentials, as for example where 
the de Vaucouleurs mass dominates, 
the temperature gradients determined by our gas dynamical model 
are slightly negative 
at most radii.
This negative gradient is due to the 
work done by gravitational compression; as gas 
expelled from stars at radius $r$ loses 
energy by radiation, it 
sinks deeper in the galactic potential.
For shallower potentials, however, $dT/dr$ can be positive, 
but such potentials are not realistic models for galaxies 
like NGC 4472.
We have not illustrated the velocity field corresponding 
to the model in Figure 2 since it is not directly observed.
The flow velocity is negative throughout most 
of the galaxy but becomes slightly 
positive in the very tenuous gas at $r \gta 160$ kpc.
Within this radius the velocity gradually increases inward
reaching -20, -50, and -140  km s$^{-1}$ at
10, 1, and 0.32 kpc respectively.

The mean interstellar iron abundance in model STD, weighted by 
$\rho^2 \Lambda$, is
$\langle z_{Fe} \rangle = 1.03$ 
in solar units which is higher than the mean 
(mass-weighted) 
stellar value $z_{* Fe} \approx 0.34$.
If the observed interstellar abundance 
in NGC 4472 is as low as Arimoto et al. (1997) claim, 
$z_{Fe} \approx 0.33$, then our (rather low) current 
supernova rate 
SNu$(t_n) = 0.04$ may still be an overestimate.

In model STD 
essentially all of the gas lost from the stars is retained 
within the galaxy.
Most of the gas created in model STD from $t = 1$ to 13 Gyr,
about $7.2 \times 10^{10}$ $M_{\odot}$, 
eventually cools at the very center of the flow 
leaving only a relatively 
small mass of hot gas, $0.6 \times 10^{10}$ $M_{\odot}$, to 
produce the x-ray luminosity $L_x$.
In more realistic rotating cooling flows the gas cools to 
a disk (Brighenti \& Mathews 1996; 1997b).
In Table 1 
we list several global parameters of the 
standard model STD and its variants described below.

\section{VARIATIONS ON THE STAN\-DARD MODEL}

\subsection{Mass Dropout: Model DO}

The conventional theoretical 
means of flattening steep density 
profiles in galactic cooling flow models 
is straightforward: remove hot gas from the 
cooling flow near the galactic center!
In addition to correcting the density profile,
this much-used assumption of ``mass dropout'' was 
also motivated by 
the possibility that thermal instabilities could 
develop from 
hypothetical inhomogeneities in the cooling flow gas
(Fabian \& Nulsen 1977; Stewart et al. 1984;
White \& Sarazin 1987a, 1987b, 1988; 
Thomas et al. 1987; Vedder, Trester \& Canizares 1988;
Sarazin \& Ashe 1989; Bertin \& Toniazzo 1995).
However, such thermal instabilities do not grow in the 
expected manner.
Low amplitude thermal instabilities are 
generally stable (Balbus 1991)
and coherent non-linear perturbations of larger amplitude 
oscillate radially in the cooling flow atmosphere 
(Loewenstein 1989) so that they are alternatively overdense 
and underdense relative to ambient gas, 
reducing or eliminating the growth rate of the instability.
Even if instability were possible, 
such initially overdense, oscillating regions
are not likely to remain coherent since hydrodynamical 
models show conclusively 
that they disintegrate by Rayleigh-Taylor 
and Kelvin-Helmholtz instabilities before completing 
a single oscillation (Hattori \& Habe 1990; Malagoli et al. 1990).
Moreover, 
the origin of the required inhomogeneities is also unclear
since there is no known source of 
entropy perturbations on the appropriate scales 
in the interstellar gas (Mathews 1990). 

Nevertheless, the problem of steep theoretical density profiles
is clearly still present in Figure 2!
In view of this and 
in deference to those many astronomers who 
have invoked mass dropout, we now describe some models 
for NGC 4472 using the standard dropout assumptions 
which are expected to flatten $d \rho / d r$.
When dropout is included, an additional term 
$-q {\dot \rho} / t_{do}$ is introduced on the right hand side 
of the equation of continuity but the equations of motion 
and thermal energy remain unchanged.
The characteristic dropout time is usually assumed to be the 
time for local radiative cooling,
$t_{do} = 5 m_p k T / 2 \mu \rho \Lambda$ 
(e. g. Sarazin \& Ashe 1989), and the constant
dimensionless parameter $q$ controls the amount of mass 
dropout.
Gas is removed from the flow on the spot, i.e. $t_{do}$
is assumed to be much less than the local flow time.
Although $t_{do} \ll t_{flow} \approx r/|u|$
is widely used in mass dropout calculations,
it is often strongly violated. 
The rate that gas is removed from the flow is proportional to 
$\rho t_{do}^{-1} \propto \rho^2$
and therefore 
increases closer to the galactic center as required 
to flatten $d \rho / d r$.
The temperature of the {\it remaining} gas that does not 
participate in the dropout is increased.
This occurs since 
after dropout less gas is present at small radii to support the 
weight of the same large global mass of interstellar gas 
so compressional
heating is intensified at small radii 
to maintain hydrostatic pressure equilibrium.
However, as gas cools and 
undergoes local dropout it will emit radiation 
from systematically cooler gas.
The net temperature observed at any 
radius is therefore 
a weighted mean between the background flow and the 
gas that is dropping out.
The local x-ray emissivity $\epsilon_{\Delta E}$ 
into the ROSAT band ($\Delta E = 0.2 - 2.4$ keV) 
is increased by a factor $(1 + q \Delta_0)$ and the 
local mean temperature is lowered 
from $T$ to $T_{eff} = T(1 + q \Delta_1)
/(1 + q \Delta_0)$ where $T$ is the temperature of the 
gas that has not yet dropped out and $\Delta_0$ and 
$\Delta_1$ are defined by 
$$\Delta_n = {1 \over T^{1+n}}~
{\Lambda \over \Lambda_{\Delta E}}~
\int_0^T (T')^n {\Lambda_{\Delta E} \over \Lambda} dT'.$$
The functions $\Delta_0(T)$ and $\Delta_1(T)$ are plotted 
in Figure 3.

Figure 4 shows the density and temperature structure of 
a cooling flow with mass dropout using $q = 1.2$.
A comparison with Figure 2 reveals that 
$d \rho / d r $ is indeed flattened within about 40 kpc 
by the dropout but
the flow at larger radii, where the dropout is very small, 
is unchanged.
The slope $d \rho / d r $ between 1 and 10 kpc is almost
flattened to the slope of the NGC 4472 density profile 
and a further increase 
in $q$ would bring the slope there into agreement.
However, dropout has also {\it lowered} 
the gas density in the bright
inner galaxy so $\alpha_*(t_n)$ would need to be increased 
by $\gta 4$ to bring the dropout solution into approximate 
agreement 
with local NGC 4472 densities within 10 kpc.
Such a large increase in $\alpha_*(t_n)$ may be inconsistent
with reasonable stellar IMF slopes and mass cutoffs
(Mathews 1989).
In any case, the gas density is very much lower than observed
values at $r \sim 100$ kpc where the x-ray observations are 
still very reliable. 
Although the beneficial influence of dropout on the density 
profile near the core has long been recognized, its 
failure to flatten $d \rho /d r$ at larger radii has been 
unappreciated probably because most previous dropout solutions
also used the steady state approximation which is known to diverge
and become unreliable 
at large galactic radii (e.g. Vedder, Trester \& Canizares 1988).

But the most serious problem for the cooling flow with dropout 
is seen in the temperature profile in Figure 4. 
The light solid line shows the temperature variation of the 
background gas which is increased over that in Figure 2 as 
explained above.
When the cooler temperatures of locally dropping out gas is 
included, however, the apparent gas temperature 
$T_{eff}(r)$ (heaver solid line in Figure 4) 
is slightly {\it lower} than the temperature without
dropout shown in Figure 2.
According to the usual dropout assumptions, 
the dropout causes a lowering of the gas temperature 
even at very large galactic 
radii (Figure 4) 
since the small mass that drops out there is compensated 
by the much longer dropout time $t_{do}$.
Therefore, even if mass dropout were a physically acceptable
process, it fails to correct the deficiency of gas at large 
galactic radii (large negative $d  \rho / d r$) 
and moves the computed gas temperatures further 
from observed values.

Aside from the many theoretical arguments against the widely-used 
assumption of mass dropout, x-ray observations also indicate that
galactic cooling flows cannot have a pronounced 
global multiphase character 
in which regions of low entropy are cooling out.
Unless there is a conspiracy of compensating effects, the 
near perfect agreement of the total and stellar mass in 
Figure 1b -- and also for NGC 4649 (Brighenti \& Mathews 1997a) -- 
are inconsistent with mass dropout.
In the presence of dropout 
the apparent temperature is lowered from $T$ to $T_{eff}$ 
which reduces $M_{tot}(r)$ determined from the equation of
hydrostatic equilibrium. 
Also the density gradient is flatter,
as in Figure 4, implying a smaller confining 
total mass than actually exists.
But the density shown in Figure 4 represents that of the 
background flow and has not been corrected for emission 
from regions of higher density that are dropping out;
so the reduction in $M_{tot}(r)$ is not as large as would 
appear from Figure 4.
If mass dropout is really present and not allowed for 
in determining $M_{tot}(r)$ from the equation of 
hydrostatic equilibrium, 
Gunn \& Thomas (1996) have shown that
the apparent total confining 
mass is lowered by 20 to 60 percent.
In typical dropout models, 
gas is removed from the flow instantaneously.
But the pressure gradient in the flow readjusts only after a 
local sound crossing time, too slow to 
follow the effects of dropout on the gas density and temperature.
However, 
the total mass (Figure 1b) 
determined from observed gas $\rho(r)$ and $T(r)$ (Figure 1a)
agrees to within 10 percent of the independently determined 
stellar mass for $0.1 r_e \lta r \lta r_e$.
This is strong observational evidence that 
distributed mass dropout is 
not dominating the cooling flows observed
in these two well-observed ellipticals.

\subsection{Non-uniform $\alpha_*$: Model NUA}

Since mass dropout has failed to account for the observed 
interstellar gas in NGC 4472, we abandon caution and explore models 
in which the fundamental coefficients in the source terms 
of the gas dynamical equations are varied in an 
arbitrary fashion.
For example, suppose proportionally more gas is ejected from 
stars further from the galactic core, then it might be expected 
that the ISM density profile would flatten.
In an attempt to investigate this possibility,
we impose an arbitrary spatial dependence 
on the stellar mass loss coefficient in which the 
stellar mass loss is enhanced at larger radii:
$\alpha_* = \alpha_*(t_n)~(r/r_{a})^a~(t/t_n)^{-1.3}.$ 
We have considered various values of $a > 0$ and $r_a$.
A representative example of the resulting flow at $t_n$ with 
$a = 0.4$ and $r_{a} = r_e$ is shown in Figure 5.
Even though more (less) gas is being expelled from the stars 
at large (small) radii, the final solution 
for both $n(r)$ and $T(r)$ is almost identical
to that in Figure 2 with uniform $\alpha_*$!
The overall vertical normalization has changed slightly 
since the total mass lost from the stars $M_{*t}~\int \alpha_* dt$
is somewhat different, but the slope of the final density profile
is very insensitive to the radial dependence of the stellar mass 
loss rate for all of the $a$-$r_a$ pairs we have considered.
Evidently this result is due to the insensitivity of 
the gas temperature profile to $\alpha_*$ and 
the necessity for the gas to adjust to just that radial 
density distribution that achieves hydrostatic equilibrium.

\subsection{Non-uniform $\Lambda$: Model NUL}

Since none of the previous modified solutions has increased the 
gas temperature toward observed values, particularly 
further out in the galaxy, perhaps an adjustment 
of the radiative cooling rate can accomplish this.
Although 
the total radiative cooling coefficient $\Lambda(T)$ changes
slowly with temperature, $\Lambda(T = 10^6) \approx 3 
\Lambda(T = 10^7)$, an additional dependence on metallicity 
should also be present at some level and this has not been
considered in the models previously discussed.
Since the metallicity $z$ of the hot cooling flow gas 
increases inwards and $d \Lambda /d z > 0$,
we consider the evolution of cooling flows 
with arbitrarily enhanced cooling at small radii:
$\Lambda(T,r) = (r/r_s)^s~\Lambda(T)$ 
where $s < 0$ and 
$\Lambda(T)$ is the cooling rate used in the standard solution
(Figure 2).
We have solved the cooling flow hydrodynamics with 
several $s$-$r_s$ pairs (keeping all other parameters 
identical to those of the standard model) and the results for 
$s = -0.6$ and $r_s = 3r_e$ are shown in Figure 6 at time 
$t_n = 13$ Gyr.

As a result of this rather large adjustment in 
$\Lambda$ the central gas temperature has increased
slightly at $r \lta 3 r_e \approx 26$ kpc, but 
the gas temperature is still much lower than
that observed in NGC 4472.
However, at $r \gta 3 r_e$ the computed temperature 
$T(r)$ is almost unchanged from that of the standard model 
in Figure 2.
The slope of the 
density profile $d n /d r$ in Figure 6 is significantly improved 
within about 10 kpc, but is still much too small further out.
Our results for other $s$-$r_s$ pairs are similar
so we must conclude that drastic, 
unphysical adjustments to
the radiative cooling rate cannot bring the model temperatures 
(or densities) into satisfactory agreement with observations.

\subsection{Higher Supernova Rate: Models SN1 and SN2}

We now explore the possibility of raising the gas temperature 
by increasing the supernova rate above that used in the 
standard solution of Figure 2.
If the gas can be heated in this manner its pressure and density 
scale heights will also increase, flattening 
$d n /dr$ and improving solutions for 
both $T(r)$ and $n(r)$.
Of course as the supernova rate SNu$(t)$ is increased,
we can anticipate that 
the iron abundance in the hot gas will also
increase even further 
beyond the observed mean value in the hot gas
(Loewenstein \& Mathews 1991).
But we shall avert our concern about the iron abundance for the 
time being, hoping that a separate solution for reconciling 
the iron abundance can be found later by some exotic 
theoretical artifice.

The supernova rate ${\rm SNu}(t) = {\rm SNu}(t_n) (t / t_n)^{-p}$
(where $p > 0$)
can be altered either by raising $p$ to increase the past 
rate of Type Ia supernovae or by increasing the coefficient
${\rm SNu}(t_n)$ which increases SNu at all times.
Our value $p = 1$ for the standard solution 
allows for some reasonable 
cosmic evolution but is less than the exponent 
1.3 in the stellar mass loss rate $\alpha_* \propto t^{-1.3}$.
Ciotti et al. (1991) have shown that galactic winds can occur during 
the early evolution if $p > 1.3$ and propose 
that the dynamically unsteady transition from winds 
to cooling flows at the current time may 
account for the strong spread observed in $L_x/L_B$.
However, the standard solution in Figure 2 indicates a 
deficiency of interstellar gas which would be further 
reduced by galactic winds.
Therefore,
we explore here the heating effects on the 
interstellar gas using 
larger ${\rm SNu}(t_n)$ which
increases the supernova rate
equally at all times while maintaing a cooling flow 
throughout the galactic evolution.

The solid lines in 
Figure 7 show the status of the hot interstellar gas 
in our model SN1 for NGC 4472 using ${\rm SNu}(t_n) = 0.25$, about 
six times larger than the standard model in Figure 2.
The gas temperature throughout the galaxy 
has in fact been increased, 
but only by about 20 percent, far less than the factor 
that ${\rm SNu}(t_n)$ is increased.
The gas density profile in Figure 7 is also significantly improved,
but is still steeper than the observed slope.
As expected, the mean iron abundance 
in the interstellar gas $\langle z_{Fe} \rangle \approx 4.41$ 
(Table 1) far exceeds any of the
values observed in NGC 4472 and 
by a much larger factor than the standard model STD.

Although increasing the supernova rate has a beneficial influence 
on both $n(r)$ and $T(r)$, ${\rm SNu}(t_n)$ cannot be raised 
further without driving a galactic wind. 
Already when ${\rm SNu}(t_n) = 0.3$ much of the interstellar 
medium is moving outwards at time $t_n = 13$ Gyrs and the 
gas density has declined substantially throughout the galaxy.
The dashed contours in 
Figure 7 show the density and temperature structure 
at $t_n$ for the wind solution SN2 
that develops when ${\rm SNu}(t_n) = 0.45$.
The density has dropped to unacceptably low values 
and the temperature far exceeds observed values.

\section{MODELS WITH ADDITIONAL CIRCUMGALACTIC GAS: 
MODELS EXG1 AND EXG2}

Having considered a variety of both modest 
and radical perturbations 
of our initial standard model for NGC 4472,
all the same problems still remain: the slope of the density 
is too steep, the mass of hot gas at large galactic radii 
is too small, 
the temperature is too low by about a factor of 2, and 
the shape of the temperature profile is wrong.

A clue to the resolution of this problem may lie in the 
total amount of hot gas observed in NGC 4472
(Trinchieri, Fabbiano, \& Canizares 1986; 
Irwin \& Sarazin 1996; Brighenti \& Mathews 1997a).
At $r = 140$ kpc, the largest galactic radius at which reliable 
gas density 
measurements are available for NGC 4472, 
the total mass in hot gas is about 
$M_{gas} = 3.5 \times 10^{10}$ $M_{\odot}$, 
about 7 percent of the total stellar mass of the galaxy. 
Using our standard mass loss rate $\alpha_*(t)$ 
we expect that approximately 
8.5 percent of the total stellar mass of the galaxy should have been
ejected into the interstellar medium from 1 to 13 Gyrs.
However, allowance must be made for the large fraction of 
this gas that cools. 
The mass budget of our standard 
model in Table 1 indicates 
that about 92 percent of the gas 
ejected by stars over this cosmic time interval has already 
cooled and is no longer available to emit x-rays.
Therefore, by analogy with the standard solution,
the amount of gas observed in NGC 4472 out 
to 140 kpc implies a minimum initial total gas mass of at least 
$6.3 \times 10^{11}$ $M_{\odot}$ -- this is about 10 times 
larger than the mass that can be ejected by galactic stars!
We conclude that {\it most of the mass that contributes to 
cooling flows in galaxies like NGC 4472 has not come from 
mass lost by galactic stars since time 1 Gyr}.

When 
extra gas is introduced into the models 
many new and uncertain parameters must be determined.
Is most of the extra gas present at early times or is it slowly 
added during the Hubble time?
What is the origin of this extra gas: is it primordial or has it been 
enriched and ejected by galactic winds from the massive 
elliptical or other nearby galaxies?
What is the temperature and density (entropy) of this extra gas?
Since the additional gas must be about twice as hot as that in 
the standard solution, 
what total mass of dark matter is required to bind the additional
gas?
Although we have explored some of these questions, 
we shall not attempt to answer them in detail here.
Our more limited objective here is to 
present results of some approximate and 
preliminary calculations that nevertheless conclusively 
illustrate the many benefits of additional hot gas 
in understanding 
the nature of cooling flows in bright elliptical galaxies.

Our provisional astrophysical interpretation of the source of 
the extra gas required, as discussed below, 
is that all bright ellipticals with 
$T_{gas} \sim 1.5 - 2 T_*$ (Davis \& White 1996) 
were originally formed in small galaxy groups of the 
sort discussed by Mulchaey et al. (1996).
These authors have shown that a significant fraction of 
small groups of galaxies contain 
dominant bright ellipticals and that most or all of the 
observed diffuse, x-ray emitting gas in these groups 
is usually closely associated with the central massive 
elliptical.
This implies that the source of the hot gas is closely related 
to the creation of bright 
elliptical galaxies, probably by mergers and tidal 
mass transfers.
Dynamical studies of the evolution of galaxies in 
elliptical-forming groups 
(Merritt 1985; Bode et al. 1994; Garciagomez et al. 1996;
Athanassoula et al. 1997; Dubinski 1997) require that 
ellipticals were formed early in the dynamical evolution of 
groups when mergers and tidal interactions were 
most likely;
if so, both the elliptical galaxy and the hot interstellar gas 
surrounding it are very old.
With this in mind, we consider models in which 
the extra hot gas is already present at 
the time $t = 1$ Gyr when our 
model calculations begin.
We shall not discuss here the alternative 
possibility that gas gradually flows into large elliptical 
galaxies over the Hubble time.

To construct a variant of our standard model
having large amounts of hot gas 
at $t = 1$ Gyr, we consider a
(somewhat arbitrary) gas density distribution 
at this early time described by 
$n(r) = n_o [ 1 + (r/r_c)^2]^{-3/4}$ where $n_o = 0.5$
cm$^{-3}$ and $r_c = 0.20$ kpc, 
out to radius $r = 62$ kpc.
At $r > 62$ kpc we assume 
$n(r) = 10^{-4}$ cm$^{-3}$ out to $r = 470$ kpc where 
the gas is abruptly terminated.
The initial gas temperature $T = 1.2 \times 10^7$ K 
and iron abundance $z_{Fe} = 0.1$ are uniform 
in this extended gaseous halo.
We consider two models beginning with this
extra gas: in the first (EXG1) $\alpha_*$ 
has the standard value and time dependence; in the 
second (EXG2) $\alpha_*$ is reduced by half at all times.
All other parameters are those of our standard solution.

As in the previous calculations, we assume that the dark 
halo of our model for NGC 4472 extends out to 470 kpc where 
stars and dark halo have a combined total mass of 
$27  \times 10^{12}$ $M_{\odot}$.
This is large, but comparable with the masses of small 
groups of galaxies found by Mulchaey et al. (1996).
The total initial mass of hot gas is 
$M_{hot} = 1.50 \times 10^{12}$ $M_{\odot}$ within 470 kpc, 
but only $M_{hot} = 0.019 \times 10^{12}$ $M_{\odot}$ of this 
gas is within 100 kpc which includes the stellar part of 
the galaxy.
Since the total stellar mass is $0.72 \times 10^{12}$ $M_{\odot}$,
the initial ratio of total baryonic to dark matter is 0.08, 
acceptably close to the expected cosmological value.
It is clear from simple dynamical considerations 
that the dark matter halo must extend 
out considerably beyond ROSAT observations at 140 kpc 
since if the gas and dark matter were truncated at this radius
a rarefaction wave would move into the galaxy in $\lta 1$ Gyr 
and sharply reduce gas 
densities and temperatures at $r \lta 140$ kpc.
Therefore our assumptions about the mass of initial gas and 
dark matter are consistent with current observations of 
NGC 4472.

In Figure 8 we show temperature and density profiles in the
hot interstellar gas at $t_n = 13$ Gyrs 
for two models having the initial gas 
configuration described above.
The first model (EXG1), shown with dashed lines, 
is computed with other 
parameters identical to the standard model.
The gas density is in excellent agreement with the observations 
for radii $r \gta 6~{\rm kpc} = 0.7r_e$ but steepens further in.
In the second model (EXG2) $\alpha_*$ is reduced by a 
factor of 2 so that the galactic stars contribute less to the 
density profile within $r_e$.
This model shown with solid contours in Figure 8 
is an even better fit 
to the overall observed density profile and its luminosity,
$L_x = 49 \times 10^{40}$ ergs s$^{-1}$, is very close to 
that observed for NGC 4472,
$L_x = 45.6 \times 10^{40}$ ergs s$^{-1}$ 
(Eskridge, Fabbiano, \& Kim 1995) when scaled to $d = 17$ Mpc.
For either model, the gas density near the outermost observed 
radius ($\sim 140$ kpc) 
is in excellent agreement with observations.
The temperature profile for either model is also in adequate 
agreement with observations.
This consistency with current observations of 
NGC 4472 follows from the extended region of 
(constant density) hot gas beyond 62 kpc that we 
assume at time $t = 1$ Gyr; 
if only this component of the initial hot gas is retained, the 
agreement of the model at $t = 13$ Gyrs 
with observed $n(r)$ and $T(r)$ is about the same as 
that shown in Figure 8.
Our objective here is not to cosmetically 
adjust the parameters of the 
initial gas distribution by trial and error until the agreement 
with the present day $n(r)$ and $T(r)$ is perfect.
Instead our less ambitious objectives are 
(i) to demonstrate that very hot gas can be retained by 
massive ellipticals throughout the Hubble time and (ii) that this 
gas is the key to understanding present-day observations.
In a more comprehensive model, 
the detailed properties of the initial (or subsequently
added) gas must be a natural consequence of more global 
cosmological considerations.

\section{ASTRONOMICAL IMPLICATIONS OF CIRCUMGALACTIC GAS}

We conclude that gas ejected from stars evolving since 
the epoch of strong galactic winds,
$\sim 1$ Gyr, is insufficient to account for the density and thermal 
structure observed in elliptical cooling flows.
This conclusion is not unique to NGC 4472. 
Davis \& White (1996) have shown that ROSAT temperatures for bright 
ellipticals are systematically higher than mean stellar temperature 
by $\sim 1.5 - 2$.
Brighenti \& Mathews (1997a) have noticed that the 
normalized radial variation of 
gas temperature $T(r/r_e)$ for six bright ellipticals having good 
ROSAT and ASCA temperatures is remarkably similar,
regardless of the environmental context of the galaxy.
Some of these galaxies are dominant ellipticals in small
groups or poor clusters 
(NGC 5044: David et al. 1994; NGC 507: Kim \& Fabbiano 1995;
NGC 1399: Jones et al. 1997) while others are associated with 
the Virgo cluster (NGC 4636: Trinchieri et al. 1994; NGC 4472: Irwin
\& Sarazin 1996; NGC 4649: Trinchieri et al. 1997).
Because of the apparently universal thermal structure among 
these galaxies, we conjecture that all these bright ellipticals 
originally formed as the primary galaxy in a small group.
The uniformity in the currently observed temperature profiles 
implies that the initial evolution of these ellipticals 
must have been very similar.
With time some of these luminous ellipticals 
have now entered larger clusters 
where they have retained much of the gas inherited from the group
environment.
Certainly 
NGC 4472 is consistent with this interpretation
since it is the central galaxy in a Virgo sub-cluster.
Although all three Virgo galaxies (NGC 4636, 4472, and 4649) 
have the same positive temperature gradient 
at $r \lta 3r_e$, NGC 4649 has noticeably less hot interstellar gas 
than 4472 and 4636.
This may be explained by the apparent proximity of NGC 4649 to 
the center of the Virgo cluster: hot interstellar gas 
at large galactic radius that 
initially contributed to the production of $dT/dr > 0$ in 
the central regions of NGC 4649 has since been ablated away by 
the velocity of 4649 relative to the intercluster medium.
The pressure of the initial outer gaseous halo in 4649 has 
now been replaced by that of the diffuse hot gas in Virgo.
Alternatively, NGC 4649 could have been 
originally a subordinate member 
of a small group of galaxies when its outer halo was
tidally transferred 
to the group-dominant elliptical.
These considerations are only appropriate for massive 
ellipticals for which current observational data 
for $\rho(r)$ and $T(r)$ are available.
The radial interstellar structure in 
less luminous ellipticals may differ: they 
may have less hot gas, lower overall temperatures and and steeper 
density gradients, resembling our standard model in Figure 2.
Perhaps AXAF can answer this question in more detail.

How is the characteristic temperature profile $T(r)$
in massive galaxies produced? For further insight into this, we 
calculated the gas dynamical flow for model EXG1:0, 
a variant of model EXG1 with 
no gas contribution from 
stellar mass loss or supernovae. After 13 Gyrs 
the temperature of model EXG1:0 is in excellent agreement 
with observed temperatures of NGC 4472 
within about 8 kpc from the galactic center
(see Figure 8).
At larger radii, however, 
the gas is much hotter, rising to a maximum 
$2 \times 10^7$ K at 35 kpc then dropping to 
$1.5 \times 10^7$ K at 80 kpc.
At all galactic radii shown in Figure 8, the temperature of 
externally provided gas in our model with no 
stellar gas $T(EXG1:0)$ is
much higher than $T(STD)$, the gas temperature 
resulting from stellar mass loss alone (Figures 2 and 8).
Evidently, the reasonably successful temperature agreement 
of model
EXG1 with NGC 4472 observations is due in part to the mixing 
of hot circumgalactic gas with relatively cooler gas ejected from
the stars.
At any radius $r \lta 100$ kpc (which includes all of the stellar
system),
$T(EXG1)$ is an approximate thermal average of $T(STD)$ and 
$T(EXG1:0)$, weighted by the relative masses contributed at each 
radius by stellar mass loss and inflow of circumgalactic gas.
By tagging the external gas and that from the stars with different
abundances, we have determined (at time $t = 13$ Gyrs) 
for model EXG1 that about
60-70 percent of the gas at $r \lta r_e = 8.57$ kpc 
comes from the stars  
while at $r \approx 75$ kpc the stellar 
contribution has dropped to 10 percent. 
This fractional contribution is consistent with
the relative temperatures of the three models
shown in Figure 8 where it is seen that 
$T(EXG1:0) > T(EXG1) > T(STD)$ at all radii,
although $T(EXG1:0)$ is closer to $T(EXG1)$ at large radii
and closer to $T(STD)$ at $r \lta r_e$.
We recognize that this argument is only approximate 
because the flow structure
$u(r)$ in these three models is different.
Nevertheless, it is clear that the characteristic 
temperature profile $T(r)$ observed in many bright ellipticals
including NGC 4472 
-- rising from $T \sim T_*$ near the origin to 1.5 - 2$~T_*$ 
at $r \sim 3r_e$ then becoming approximately isothermal --
is a signature of the thermal combination of 
two sources of hot interstellar gas.

Where does the circumgalactic gas come from and why is 
it so hot?
At present we cannot be entirely certain about the origin of this
gas, but it is likely that its original source was 
``secondary infall'' or the reversal of the Hubble flow 
surrounding positive density fluctuations that grow to become 
massive ellipticals or small groups.
Upon arrival at the young galaxy, the 
infalling gas passes through a strong shock
that raises its temperature to approximately the local virial
temperature $T_{vir} \approx (\mu m_p/2 k)GM_{tot}/r$ in the halo
(Bertschinger 1985).
It is likely that Type II supernovae from massive stars
further heated and enriched the interstellar gas 
during the early stages of galactic formation.
In the mass distribution for NGC 4472 (Figure 1b) 
$T_{vir}$ increases by a factor 2.1 from $r = r_e$ to 
$r = 10r_e$; this may explain why the circumgalactic
gas located at $r \gg r_e$ is hotter than the stars 
as required in our successful models and indicated by 
the observations of Davis \& White (1996).
Subsequently, the hot halo gas and dark matter may have been  
dynamically transferred between galaxies while they were 
members of small groups.

The introduction of rather large masses of 
hot gas into ellipticals at early times may 
either resolve or complicate 
other problems that have received much attention over the years.
For example, the wide scatter in $L_x$ for given optical luminosity 
$L_B$ may result from tidal exchanges of this 
extended halo gas component between ellipticals
having similar $L_B$ within the small groups in which they were formed
(Mathews \& Brighenti 1997).
The general circularity of x-ray isophotes in rotating, massive ellipticals 
must also be understood in the context of circumgalactic gas. 
Brighenti \& Mathews (1996) have shown that even the modest rotation 
of the stellar component observed in 
massive ellipticals is sufficient to produce strongly rotating disk 
configurations in the hot interstellar gas, assuming that 
the interstellar gas derives only from stellar mass loss.
The x-ray images are flattened perpendicular to the axis of rotation.
However, if the gas expelled from stars is overwhelmed by inflowing 
low angular momentum hot gas from the outer galaxy, the x-ray
isophotes will appear more circular.
But it is unclear if the circumgalactic gas would be expected 
to have less rotation.
[Flattened x-ray isophotes may nevertheless still be present in 
rapidly rotating 
ellipticals of moderate or low luminosity (Brighenti \& Mathews 1997b)].
The presence of additional gas in bright ellipticals 
both complicates
and clarifies the interpretation of the iron abundance observed in 
the hot gas.
Even when no circumgalactic  
gas is present, gas dynamical cooling flow models predict 
that the interstellar iron abundance should be less than that in 
the stars, provided the Type Ia supernova rate is very low; this occurs 
because the hot gas flows inward through 
a stellar system having a negative radial abundance gradient.
But the disparity
between high stellar and low gas abundances 
is so large that some have questioned the validity of 
abundance determinations from x-ray spectra 
(Arimoto et al. 1997; Renzini 1997). 
The Type Ia supernova rate in our standard model STD 
is comparable with the low rate observed,
but still may produce too much iron.
Nevertheless, we expect that observed iron abundances 
represent a combination of iron 
from stellar ejecta and circumstellar gas -- of unknown 
abundance -- that flows in from large galactic radii.
The iron abundance in our model EXG2 is in fact lower than that
of the standard model (Table 1), but not by very much 
for the parameters we considered.
However, Loewenstein (1996) has pointed out 
that ellipticals having the largest amount of 
extended, circumgalactic gas have on average {\it higher}
iron abundances, contrary to the expected trend.

\section{CONCLUSIONS}

We have attempted without success 
to derive the known density and temperature 
structure in NGC 4472 using the usual equations of gas dynamics 
and standard cooling flow assumptions. 
Our mass model for NGC 4472,
based on hydrostatic equilibrium, is fully self-consistent
and is not likely to be incorrect.
In our attempts to bring the results of 
dynamical models into agreement 
with observations of NGC 4472, we considered mass dropout,
spatially variable stellar mass loss and radiative 
emission rates,
and higher supernova rates.
Invariably, 
the total amount of hot 
gas generated by stellar mass loss that remains at the 
present time 
is insufficient, particularly 
at large galactic radii, and the density slope  
is too steep.
The steepness of $d \rho /d r$ is obviously 
correlated with the relatively low gas temperatures 
$T \approx T_*$ 
that result 
from thermalization of stellar mass loss and 
supernova rates 
limited by the known iron abundance in the interstellar gas.

All these difficulties are resolved when additional hot gas is 
provided from the outer regions of the galactic potential.
We are not the first to consider additional gas flowing into 
galactic cooling flows.
Within the context of steady state solutions, both 
Thomas (1986) and Bertin \& Toniazzo (1995) 
discussed the advantages of inflowing gas at the 
outer parts of the galaxy and both studies constrained
their theoretical results to match observed properties 
of NGC 4472.
The need for large gas masses
in addition to that provided by stellar mass loss
is already evident from the 
high gas mass indicated by ROSAT observations of 
NGC 4472 and other similar bright ellipticals.
The gas density in NGC 4472 
at $\sim 140$ kpc, the outer 
limit of reliable x-ray surface brightness observations, 
is unstable to rapid outflow unless 
both the gas and the dark halo extend to significantly larger radii.
We have therefore considered an 
initial configuration for NGC 4472 in which most of 
the original baryonic matter is in 
an extended hot gas phase and have 
shown that much of this gas can be retained over the Hubble time.
Our model for the initial gas configuration is by no means 
unique, but serves to demonstrate the possible retention of 
very old hot gas in massive ellipticals.

The total mass of initial hot gas we consider
within the large galactic halo 
(of radius $r = 470$ kpc) at $t = 1$ Gyr,
$M_{hot} = 1.50 \times 10^{12}$ $M_{\odot}$,
is about 2.1 times larger than the stellar mass.
The temperature is initially 
uniform and hot, $T = 1.2 \times 10^7$ K. 
This gas and the dark galactic halo are assumed to  
extend out to $r = 470$ kpc.
After following the gas dynamical evolution, 
by time 
$t_n = 13$ Gyr the mass of hot gas within $r = 470$ kpc 
is reduced by cooling and outflow to 
$0.40 \times 10^{12}$ $M_{\odot}$.
The mass of hot and cooled gas within the central, stellar part 
of the galaxy, $r \lta 100$ kpc, 
$M_{hot} = 0.04 \times 10^{12}$ $M_{\odot}$
and $M_{cold} = 0.03 \times 10^{12}$ $M_{\odot}$,
are both very much less than the total stellar mass,
$M_{*t} = 0.72 \times 10^{12}$ $M_{\odot}$.
If the cooled gas formed into relatively low mass stars, the 
stellar mass to light ratio would not be affected.
As gas from the 
reservoir of hot gas in the outer galaxy 
flows inward toward the stellar component,
its temperature remains high,  
similar to values observed, $T \sim 1.5 - 2 T_*$ and 
positive temperature gradients are generated within a few $r_e$,
just like the observed pattern.
Although both gas and stars reside in the same potential, 
the stars are confined to just the central regions and therefore 
have a mean temperature that is less than the more extensive gas.
Since the gas is hotter, the gas density profile is flattened
at all radii in conformity with observation.

The addition of hot gas to the outer regions of elliptical 
galaxies, either 
initially or over time, does not account only for 
the observations of NGC 4472,
but is essential to understand similar 
model deficiencies in all massive ellipticals.
Moreover, the similarity in observed $T(r)$ 
in most or all bright ellipticals  
observed so far (Brighenti \& Mathews 1997a) 
-- both in dominant group galaxies 
and in members of rich clusters -- strongly 
suggests that 
all large ellipticals have retained part of the hot 
circumgalactic gas from their previous small group environment.
The elliptical-dominated groups 
observed by Mulchaey et al. (1996) may be typical of the 
early development of most or all giant ellipticals.
Some gas dynamical memory of these massive gaseous halos
persists even after bright ellipticals enter cluster 
environments.
As a bright galaxy moves deep into a large cluster, 
the pressure of hot gas in its 
outer atmosphere may be supplanted by the pressure of the 
hot diffuse cluster gas.
In this manner the characteristic 
high temperature and positive 
$dT/dr$ that are established within a few $r_e$ may 
not be greatly altered even when gas in the outer 
galaxy is stripped away; NGC 4649, deep in Virgo,
may be an example of this.

In the future it will be necessary to show that 
massive, extended halos of hot gas, similar 
to the one we require here, are an inevitable 
consequence of early galactic evolution.
The prevailing consensus today is that a significant 
outflow of gas occurred at early times during galactic 
evolution, $t \lta 1$ Gyr.
The high metallicity observed in hot diffuse cluster 
gas indicates that winds were indeed 
common in the earliest times,
either from the large ellipticals 
or from the hierarchical subunits
that later merged to form them.
It is therefore unclear how to reconcile the existence of 
the large halo of hot gas we hypothesize here with 
these early galactic winds. 
The hot gas is likely to have been acquired by massive 
galaxies over a few or many Gyrs, 
not provided all at once as we have assumed.
However, our initial investigations of models 
in which additional gas is supplied gradually 
over the Hubble time have not been encouraging; 
in general these models are much more complicated and less 
successful than the one we discussed here.

But none of these complications should 
detract from our main conclusion: 
inflowing circumgalactic 
gas may help resolve many difficulties
that have plagued the subject of cooling flows 
in elliptical galaxies.
The radial variation of gas temperature and density can 
now be understood.
The long-standing problem of the enormous scatter 
in the $L_x/L_B$ plot
(e.g. Eskridge et al. 1995) 
may result from disproportionate allocations of hot halo 
gas in group ellipticals of similar $L_B$ 
following tidal transfers of halo 
material from subordinate to group-dominant ellipticals
(Mathews \& Brighenti 1997).
We may also understand why the ratio of iron abundance in the gas to
that in the stars is so variable and often so small.
Differing amounts of incoming 
circumgalactic hot gas, assumed to have low iron abundance, 
can dilute the iron produced by galactic 
stars and supernovae in varying amounts, leading to a variety of 
(often very low) hot gas iron abundances for galaxies of similar
$L_B$.
Unfortunately, this simple idea is not supported by the few 
observations that currently exist.
Finally, the apparent 
circularity of x-ray images of large, rotating ellipticals
may be difficult to understand without additional sources of gas 
since gas lost from the stars naturally 
produces easily visible x-ray 
disks out to $\sim r_e$ even in slowly rotating ellipticals
(Brighenti \& Mathews 1996).
If circumgalactic gas flowing in from 
the outer parts of massive ellipticals has less specific 
angular momentum than the stars, the visibility
of x-ray disks would be reduced.

\acknowledgments

We wish to thank Annibale D'Ercole for allowing us to use a modified
version of his Eulerian code in performing the calculations discussed
here.  Our work on the evolution of hot gas in ellipticals is
supported by NASA grant NAG 5-3060 for which we are very grateful. In
addition WGM is supported by a UCSC Faculty Research Grant and FB is
supported in part by Grant ASI-95-RS-152 from the Agenzia Spaziale
Italiana.




\figcaption{ (a) Observations and our analytic fits
to gas temperature (top)
and density (bottom) for {\it Einstein} HRI (filled circles) and
ROSAT (open circles) data;
(b) {\it solid curve}: $M_{tot}$; 
{\it long-dashed curves}: stellar mass $M_*$ of NGC 4472;
{\it dot-dashed curves}: total mass of hot gas;
(c) {\it solid curve}: total mass density $\rho_{tot}$;
{\it long-dashed curves}: stellar mass density $\rho_*$ with a
horizontal cut at the core or break radius $r_b = 0.023r_e$.
\label{fig1}}

\figcaption{{\it top:} Filled circles and open circles are 
gas densities observed in NGC 4472 with 
{\it Einstein} HRI and ROSAT respectively. 
The line is the density variation 
determined by the standard model (Model STD) at time $t = 13$ Gyr.
{\it bottom}: Temperatures observed by ROSAT are shown with errorbars
and the radial dependence of gas temperature determined for 
the standard model at time $t = 13$ Gyr 
is shown as a solid line. \label{fig2}}

\figcaption{Plots of the dropout functions 
$\Delta_0(T)$ (solid curve) and 
$\Delta_1(T)$ (dashed curve) for the ROSAT bandpass.  \label{fig3}}

\figcaption{Variation of density and temperature in the 
cooling flow model with mass dropout (Model DO)
{\it top:} gas density;
{\it bottom}: background gas temperature of remaining 
gas (light solid line) and the mean effective temperature 
including that of gas dropping out (heavy solid line). 
The density and temperature for NGC 4472 are shown 
for comparison. \label{fig4}}

\figcaption{Variation of density and temperature in the
cooling flow model with non-uniform
$\alpha_*$ (Model NUA)
{\it top:} gas density;
{\it bottom}: gas temperature. The density and temperature for NGC
4472 are shown
for comparison. \label{fig5}}

\figcaption{Variation of density and temperature in the
cooling flow model with non-uniform
$\Lambda$ (Model NUL)
{\it top:} gas density;
{\it bottom}: gas temperature. The density and temperature for NGC
4472 are shown
for comparison. \label{fig6}}

\figcaption{Variation of density and temperature in 
cooling flow models with increased supernova rates
(Models SN1 and SN2).
{\it top:} gas density for model SN1 (SNu$(t_n) = 0.25$) 
(solid line) and for model SN2 (SNu$(t_n) = 0.45$)
(dashed line);
{\it bottom}: gas temperature for model SN1 (SNu$(t_n) = 0.25$)
(solid line) and for model SN2 (SNu$(t_n) = 0.45$)
(dashed line). The density and temperature for NGC
4472 are shown for comparison. \label{fig7}}

\figcaption{Variation of density and temperature in 
cooling flow models with additional gas at early 
times (Models EXG1 and EXG2).
{\it top:} gas density for model EXG1 (normal $\alpha_*$)
(dashed line), for model EXG2 ($\alpha_*$ reduced by 2)
(solid line), for model STD (dotted line) and for model EXG1:0 
with no stellar gas contribution (dot-dashed line);
{\it bottom}: gas temperature for model EXG1 (normal $\alpha_*$)
(dashed line), for model EXG2 ($\alpha_*$ reduced by 2)
(solid line), for model EXG2 ($\alpha_*$ reduced by 2)
(solid line), for model STD (dotted line) and for model EXG1:0
(dot-dashed line). 
The density and temperature for NGC
4472 are shown for comparison. \label{fig8}}

\clearpage

\makeatletter
\def\jnl@aj{AJ}
\ifx\revtex@jnl\jnl@aj\let\tablebreak=\nl\fi
\makeatother

\begin{deluxetable}{lcccc}
\footnotesize
\tablenum{1}
\tablewidth{0pc}
\tablecaption{GLOBAL PROPERTIES OF COOLING FLOW MODELS}
\tablehead{
\colhead{Model} &
\colhead{$L_x$\tablenotemark{a}} &
\colhead{$\langle z_{\rm Fe}/z_{\rm Fe \odot}\rangle$
\tablenotemark{b}} &
\colhead{$M_{\rm hot}$\tablenotemark{a}} &
\colhead{$M_{\rm cold}$\tablenotemark{a}} \nl
\colhead{} &
\colhead{($10^{40}$ ergs s$^{-1}$)} &
\colhead{} &
\colhead{($10^{10}$ M$\odot$)} &
\colhead{($10^{10}$ M$\odot$)}
}
\startdata
STD & 19.54 & 1.03 & 0.62 & 6.58  \cr
DO  & 12.98 & 1.15 & 0.53 & 6.7\tablenotemark{c}   \cr
NUA & 29.13 & 0.78 & 0.76 & 7.77  \cr
NUL & 8.70  & 0.97 & 0.73 & 6.38  \cr
SN1\tablenotemark{d} & 43.10 & 4.41 & 1.64 & 5.01  \cr
SN2\tablenotemark{e} & $7.66\times 10^{-4}$ & 9.64 & $1.32\times
10^{-2}$ & -- \
\
 \cr
EXG1 & 90.22 & 0.58\tablenotemark{f} & 4.62 & 7.34  \cr
EXG2\tablenotemark{g} & 49.14 & 0.85\tablenotemark{f} & 4.25 & 3.52
\cr
NGC 4472 & 45.6\tablenotemark{h} & 0.33\tablenotemark{i}
& 3.5 & - \cr

\enddata
\tablenotetext{a}{Within 100 kpc.}
\tablenotetext{b}{Averaged over x-ray emissivity.}
\tablenotetext{c}{Cold gas is distributed throughout inner galaxy.}
\tablenotetext{d}{SNu$(t_n) = 0.25$ SNu.}
\tablenotetext{e}{SNu$(t_n) = 0.45$ SNu.}
\tablenotetext{f}{Initial gas has $z_{Fe} = 0.1 z_{Fe \odot}$.}
\tablenotetext{g}{For EXG2 $\alpha_*$ is reduced by 2.}
\tablenotetext{h}{From Eskridge et al. (1995) scaled to $D = 17$ Mpc.}
\tablenotetext{i}{From Arimoto et al. (1997); see text for other
values.}
\end{deluxetable}

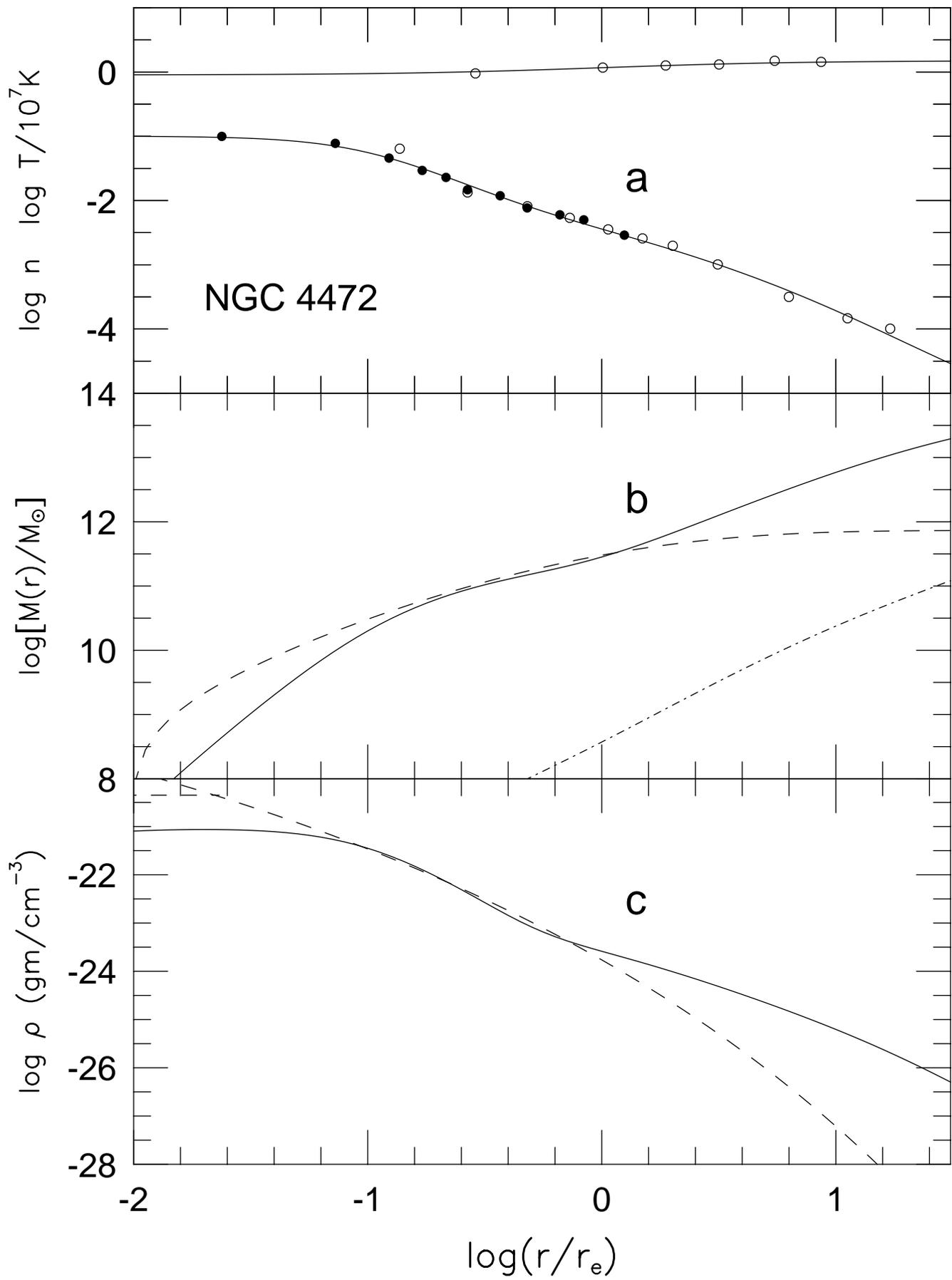
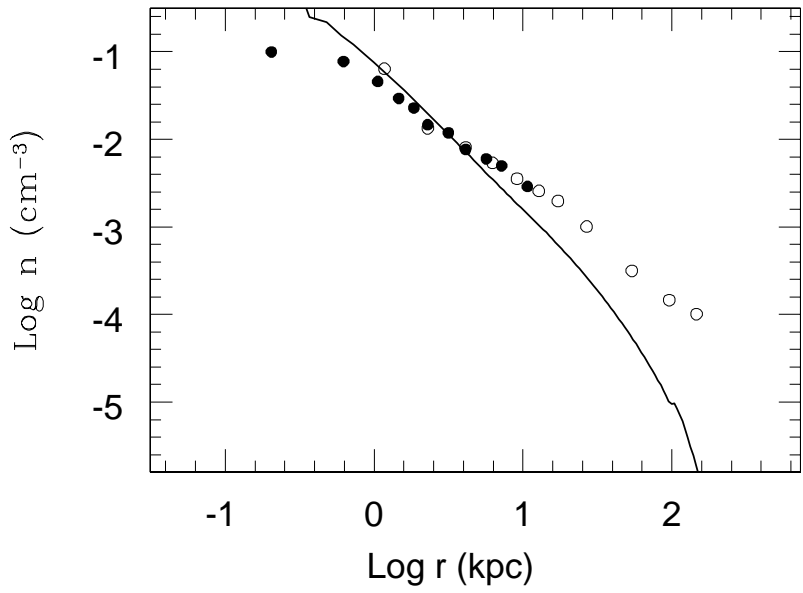
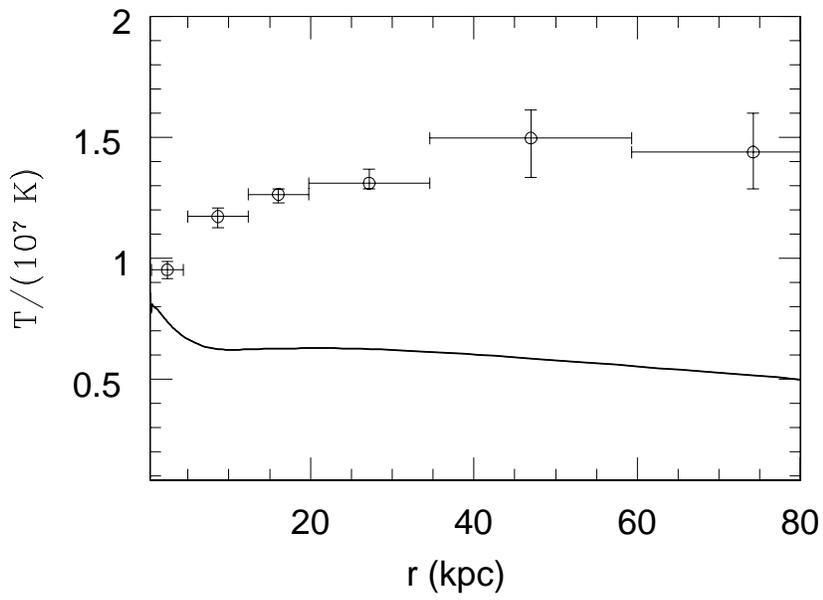
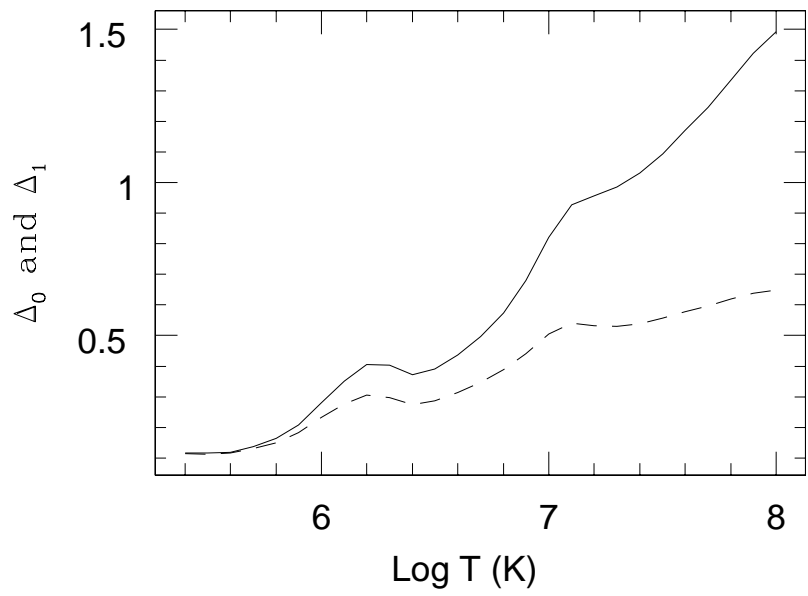
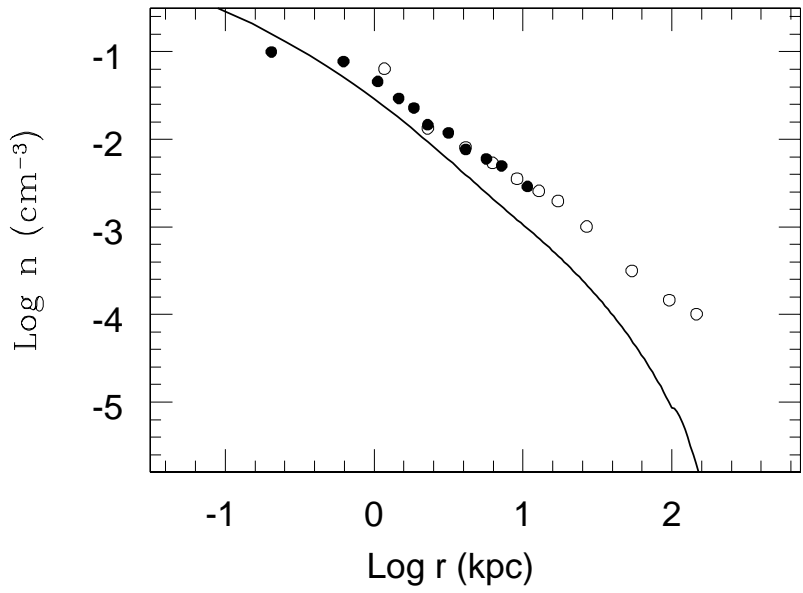
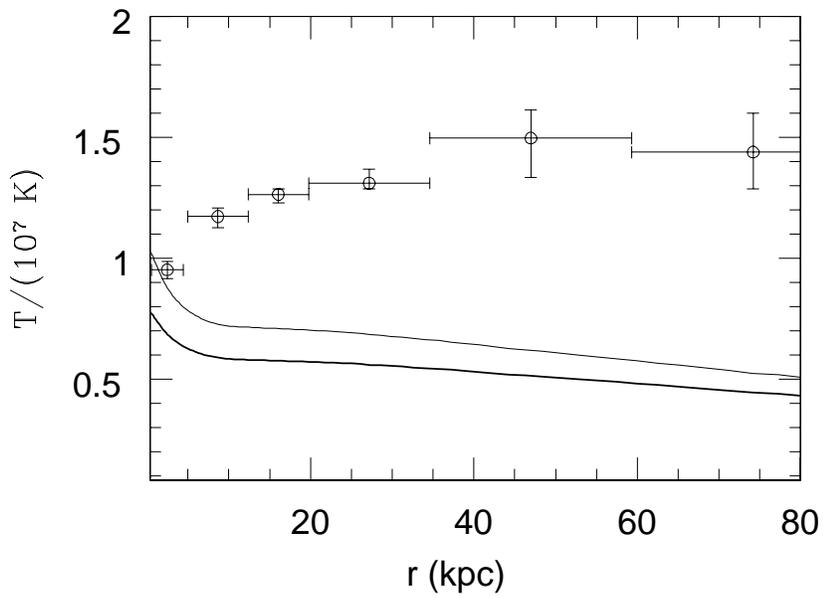
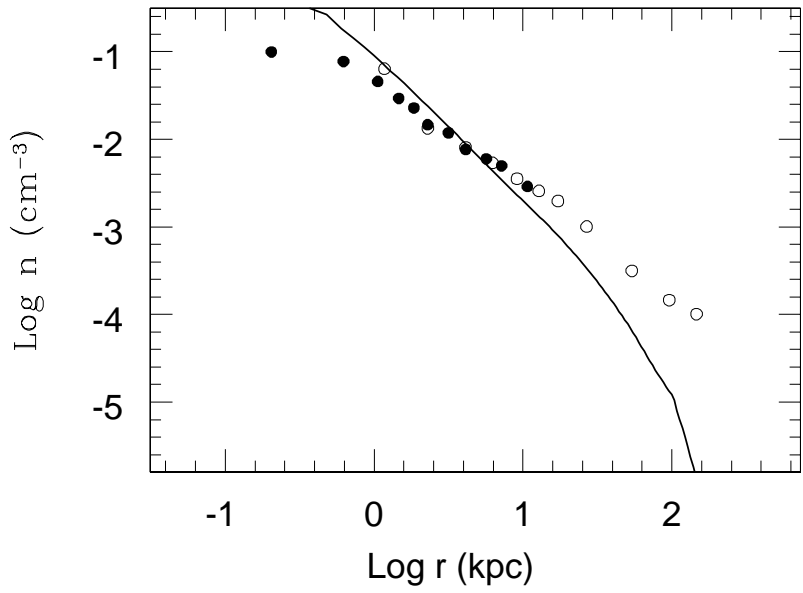
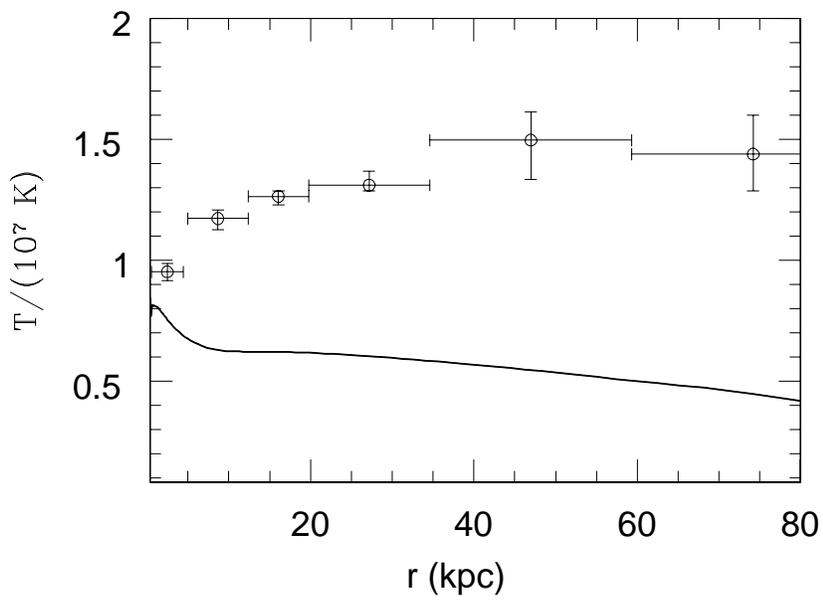
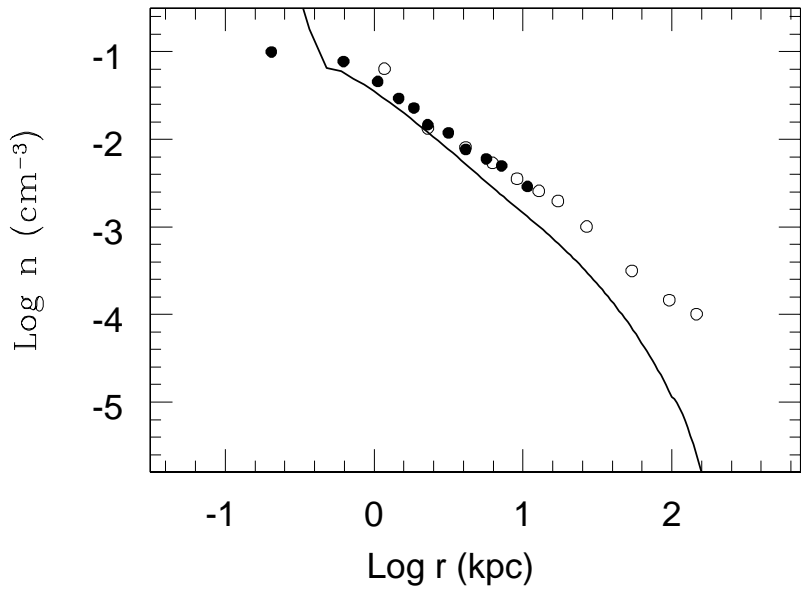
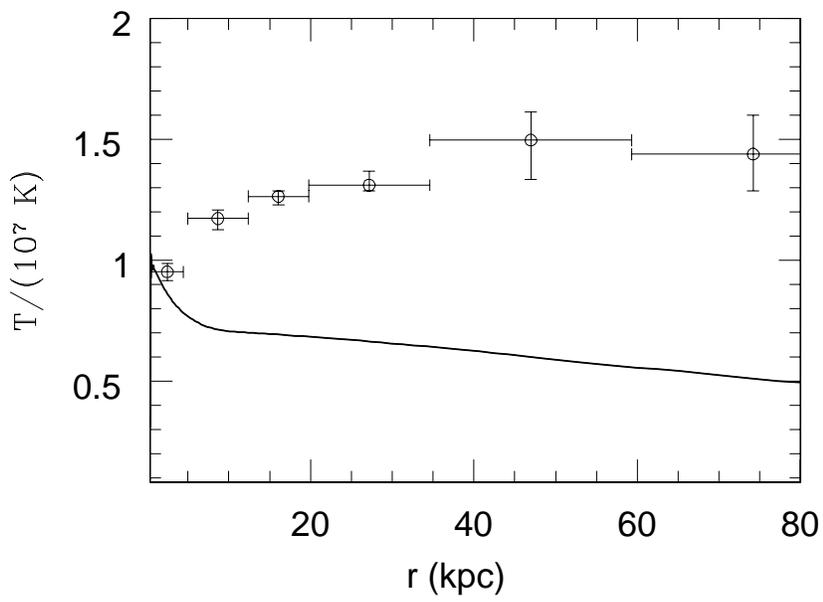
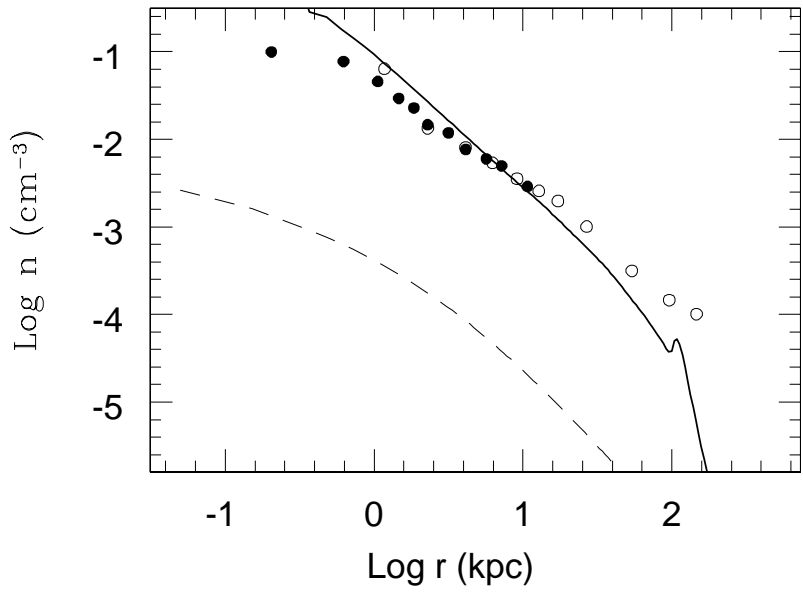
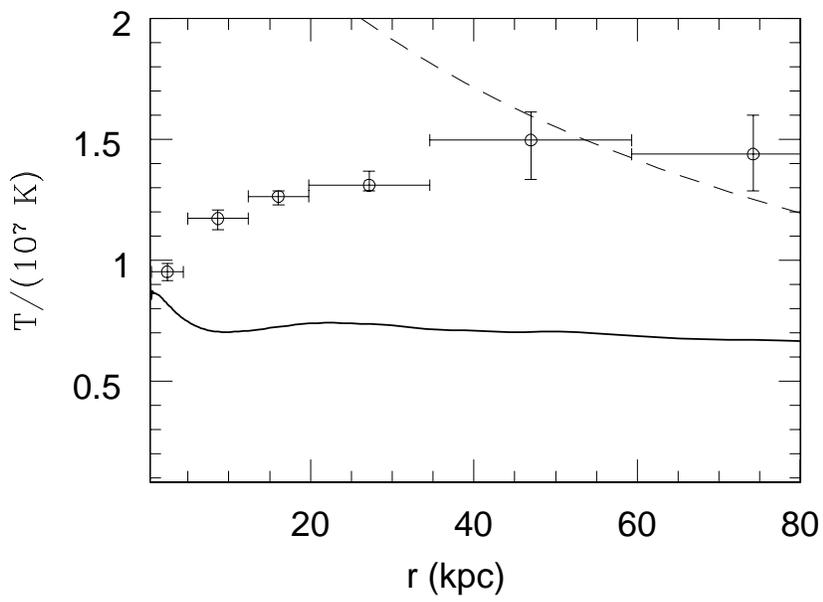
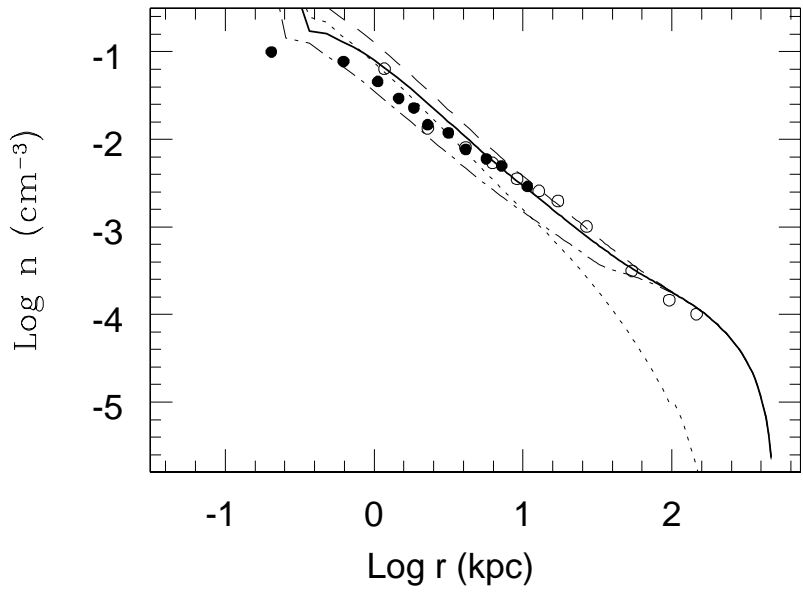
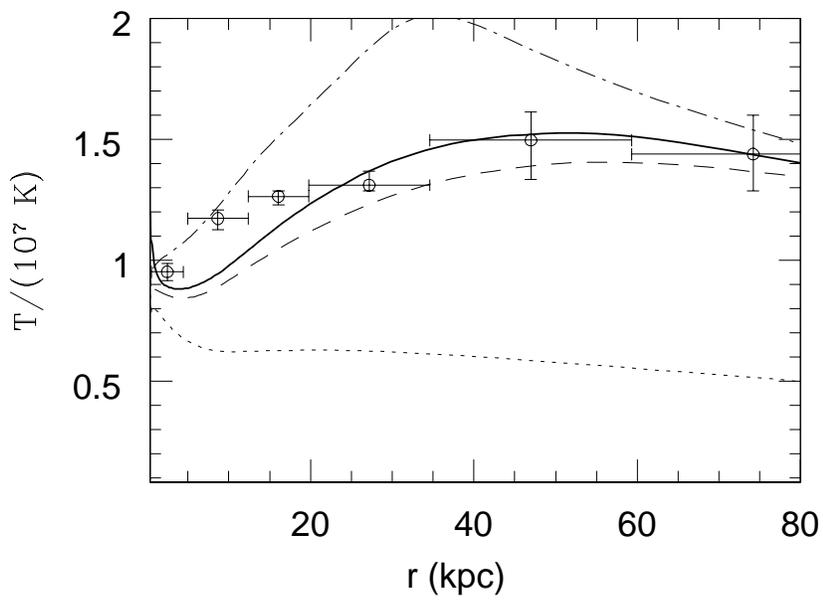
\end{document}